\documentclass{aastex7}
\usepackage{subfig}
\usepackage{booktabs}
\usepackage{listings}
\usepackage[normalem]{ulem}

\begin{document}

\title{HAWC Performance Enhanced by Machine Learning in Gamma-Hadron Separation
}

\author{R.~Alfaro}
\affiliation{Instituto de F\'{i}sica, Universidad Nacional Autónoma de México, Ciudad de Mexico, Mexico}
\email{ruben@fisica.unam.mx}

\author{C.~Alvarez}
\affiliation{Universidad Autónoma de Chiapas, Tuxtla Gutiérrez, Chiapas, México}
\email{crabpulsar@hotmail.com}

\author{A.~Andrés}
\affiliation{Instituto de Astronom\'{i}a, Universidad Nacional Autónoma de México, Ciudad de Mexico, Mexico }
\email{andres1210@gmail.com}

\author{E.~Anita-Rangel}
\affiliation{Instituto de Astronom\'{i}a, Universidad Nacional Autónoma de México, Ciudad de Mexico, Mexico }
\email{earangel@astro.unam.mx}

\author{M.~Araya}
\affiliation{Escuela de Física, Universidad de Costa Rica, Montes de Oca, San José, 11501-2060, Costa Rica}
\email{miguel.araya@ucr.ac.cr}

\author{J.C.~Arteaga-Velázquez}
\affiliation{Universidad Michoacana de San Nicolás de Hidalgo, Morelia, Mexico}
\email{juan.arteaga@umich.mx}

\author{D.~Avila Rojas}
\affiliation{Instituto de Astronom\'{i}a, Universidad Nacional Autónoma de México, Ciudad de Mexico, Mexico}
\email{doavila@astro.unam.mx}

\author{H.A.~Ayala Solares}
\affiliation{Department of Physics, Pennsylvania State University, University Park, PA, USA}
\email{hgayala@psu.edu}

\author{R.~Babu}
\affiliation{Department of Physics and Astronomy, Michigan State University, East Lansing, MI, USA}
\email{baburish@msu.edu}

\author{P.~Bangale}
\affiliation{Temple University, Department of Physics, 1925 N. 12th Street, Philadelphia, PA 19122, USA}
\email{priyadarshini.bangale@temple.edu}

\author{E.~Belmont-Moreno}
\affiliation{Instituto de F\'{i}sica, Universidad Nacional Autónoma de México, Ciudad de Mexico, Mexico}
\email{belmont@fisica.unam.mx}

\author{A.~Bernal}
\affiliation{Instituto de Astronom\'{i}a, Universidad Nacional Autónoma de México, Ciudad de Mexico, Mexico}
\email{abel@astro.unam.mx}


\author{T.~Capistrán}
\affiliation{Università degli Studi di Torino, I-10125 Torino, Italy}
\email{tcapistranc@gmail.com}

\author{A.~Carramiñana}
\affiliation{Instituto Nacional de Astrof\'{i}sica, Óptica y Electrónica, Puebla, Mexico}
\email{alberto@inaoep.mx}

\author{F.~Carreón}
\affiliation{Instituto de Astronom\'{i}a, Universidad Nacional Autónoma de México, Ciudad de Mexico, Mexico }
\email{mfcarreon@astro.unam.mx}

\author[0000-0002-6144-9122]{S.~Casanova}
\affiliation{Institute of Nuclear Physics Polish Academy of Sciences, PL-31342 IFJ-PAN, Krakow, Poland}
\email{sabrinacasanova@gmail.com}

\author{U.~Cotti}
\affiliation{Universidad Michoacana de San Nicolás de Hidalgo, Morelia, Mexico}
\email{umberto.cotti@umich.mx}

\author{E.~De la Fuente}
\affiliation{Departamento de F\'{i}sica, Centro Universitario de Ciencias Exactase Ingenierias, Universidad de Guadalajara, Guadalajara, Mexico }
\email{edfuente@gmail.com}



\author{D.~Depaoli}
\affiliation{Max-Planck Institute for Nuclear Physics, 69117 Heidelberg, Germany}
\email{davide.depaoli@mpi-hd.mpg.de}

\author{P.~Desiati}
\affiliation{Department of Physics, University of Wisconsin-Madison, Madison, WI, USA}
\email{paolo.desiati@icecube.wisc.edu}

\author{N.~Di Lalla}
\affiliation{Department of Physics, Stanford University, Stanford, CA 94305–4060, USA}
\email{niccolo.dilalla@stanford.edu}

\author{R.~Diaz Hernandez}
\affiliation{Instituto Nacional de Astrof\'{i}sica, Óptica y Electrónica, Puebla, Mexico}
\email{dihera77@gmail.com}


\author{M.A.~DuVernois}
\affiliation{Department of Physics, University of Wisconsin-Madison, Madison, WI, USA}
\email{duvernois@icecube.wisc.edu}

\author{J.C.~Díaz-Vélez}
\affiliation{Department of Physics, University of Wisconsin-Madison, Madison, WI, USA}
\email{juan.diazvelez@alumnos.udg.mx}

\author{K.~Engel}
\affiliation{Department of Physics, University of Maryland, College Park, MD, USA}
\email{kristi.engel23@gmail.com}

\author{T.~Ergin}
\affiliation{Department of Physics and Astronomy, Michigan State University, East Lansing, MI, USA}
\email{ergintul@msu.edu}

\author{C.~Espinoza}
\affiliation{Instituto de F\'{i}sica, Universidad Nacional Autónoma de México, Ciudad de Mexico, Mexico}
\email{m.catalina@fisica.unam.mx}

\author{K.L.~Fan}
\affiliation{Department of Physics, University of Maryland, College Park, MD, USA}
\email{klfan@terpmail.umd.edu}


\author{N.~Fraija}
\affiliation{Instituto de Astronom\'{i}a, Universidad Nacional Autónoma de México, Ciudad de Mexico, Mexico}
\email{nifraija@astro.unam.mx}

\author{S.~Fraija}
\affiliation{Instituto de Astronom\'{i}a, Universidad Nacional Autónoma de México, Ciudad de Mexico, Mexico}
\email{sarafraija@hotmail.com}

\author{J.A.~García-González}
\affiliation{Tecnologico de Monterrey, Escuela de Ingeniería y Ciencias, Ave. Eugenio Garza Sada 2501, Monterrey, N.L., Mexico, 64849}
\email{anteus79@tec.mx}


\author{F.~Garfias}
\affiliation{Instituto de Astronom\'{i}a, Universidad Nacional Autónoma de México, Ciudad de Mexico, Mexico}
\email{fergar@astro.unam.mx}

\author{N.~Ghosh}
\affiliation{Department of Physics, Michigan Technological University, Houghton, MI, USA }
\email{nghosh1@mtu.edu}

\author{A.~Gonzalez Muñoz}
\affiliation{Instituto de F\'{i}sica, Universidad Nacional Autónoma de México, Ciudad de Mexico, Mexico}
\email{adiv.gonzalez@itoaxaca.edu.mx}

\author{M.M.~González}
\affiliation{Instituto de Astronom\'{i}a, Universidad Nacional Autónoma de México, Ciudad de Mexico, Mexico}
\email{magda@astro.unam.mx}

\author{J.A.~González}
\affiliation{Universidad Michoacana de San Nicolás de Hidalgo, Morelia, Mexico}
\email{jose.gonzalez.c@umich.mx}

\author{J.A.~Goodman}
\affiliation{Department of Physics, University of Maryland, College Park, MD, USA}
\email{goodman@umd.edu}

\author{S.~Groetsch}
\affiliation{Department of Physics, University of Wisconsin-Madison, Madison, WI, USA}
\email{sjgroets@mtu.edu}

\author{J.P.~Harding}
\affiliation{Los Alamos National Laboratory, Los Alamos, NM, USA}
\email{jpharding@lanl.gov}

\author{S.~Hernández-Cadena}
\affiliation{Tsung-Dao Lee Institute, Shanghai Jiao Tong University, Shanghai, China}
\email{shkdna@sjtu.edu.cn}

\author{I.~Herzog}
\affiliation{Department of Physics and Astronomy, Michigan State University, East Lansing, MI, USA}
\email{herzogia@msu.edu}


\author[0000-0002-5447-1786]{D.~Huang}
\affiliation{Department of Physics, University of Maryland, College Park, MD, USA}
\email[show]{dezhih@umd.edu}


\author{P.~Hüntemeyer}
\affiliation{Department of Physics, Michigan Technological University, Houghton, MI, USA}
\email{petra@mtu.edu}

\author{A.~Iriarte}
\affiliation{Instituto de Astronom\'{i}a, Universidad Nacional Autónoma de México, Ciudad de Mexico, Mexico}
\email{airiarte@astro.unam.mx}

\author{S.~Kaufmann}
\affiliation{Universidad Politecnica de Pachuca, Pachuca, Hgo, Mexico}
\email{skaufmann13@googlemail.com}

\author{D.~Kieda}
\affiliation{Department of Physics and Astronomy, University of Utah, Salt Lake City, UT, USA}
\email{dave.kieda@utah.edu}

\author{K.~Leavitt}
\affiliation{Department of Physics, Michigan Technological University, Houghton, MI, USA}
\email{kleavitt@mtu.edu}


\author{J.~Lee}
\affiliation{University of Seoul, Seoul, South Korea}
\email{jason.lee@uos.ac.kr}

\author{H.~León Vargas}
\affiliation{Instituto de F\'{i}sica, Universidad Nacional Autónoma de México, Ciudad de Mexico, Mexico}
\email{hleonvar@fisica.unam.mx}

\author{J.T.~Linnemann}
\affiliation{Department of Physics and Astronomy, Michigan State University, East Lansing, MI, USA}
\email{linneman@msu.edu}

\author{A.L.~Longinotti}
\affiliation{Instituto de Astronom\'{i}a, Universidad Nacional Autónoma de México, Ciudad de Mexico, Mexico}
\email{alonginotti@astro.unam.mx}

\author{G.~Luis-Raya}
\affiliation{Universidad Politecnica de Pachuca, Pachuca, Hgo, Mexico}
\email{gilura6969@hotmail.com}

\author{K.~Malone}
\affiliation{Los Alamos National Laboratory, Los Alamos, NM, USA}
\email{kmalone@lanl.gov}

\author{O.~Martinez}
\affiliation{Facultad de Ciencias F\'{i}sico Matemáticas, Benemérita Universidad Autónoma de Puebla, Puebla, Mexico}
\email{omartin@fcfm.buap.mx}

\author{J.~Martínez-Castro}
\affiliation{Centro de Investigación en Computación, Instituto Politécnico Nacional, México City, México}
\email{macj@cic.ipn.mx}

\author{J.A.~Matthews}
\affiliation{Dept of Physics and Astronomy, University of New Mexico, Albuquerque, NM, USA}
\email{johnm@unm.edu}

\author{P.~Miranda-Romagnoli}
\affiliation{Universidad Autónoma del Estado de Hidalgo, Pachuca, Mexico}
\email{pa.miranda.r@gmail.com}

\author{P.E.~Mirón-Enriquez}
\affiliation{Instituto de Astronom\'{i}a, Universidad Nacional Autónoma de México, Ciudad de Mexico, Mexico }
\email{pelimi92@gmail.com}

\author{J.A.,Montes}
\affiliation{Instituto de Astronom\'{i}a, Universidad Nacional Autónoma de México, Ciudad de Mexico, Mexico}
\email{jamontes@astro.unam.mx}

\author{J.A.~Morales-Soto}
\affiliation{Universidad Michoacana de San Nicolás de Hidalgo, Morelia, Mexico}
\email{jmoralessg@gmail.com}

\author{E.~Moreno}
\affiliation{Facultad de Ciencias F\'{i}sico Matemáticas, Benemérita Universidad Autónoma de Puebla, Puebla, Mexico}
\email{emoreno@fcfm.buap.mx}


\author{M.~Najafi}
\affiliation{Department of Physics, Michigan Technological University, Houghton, MI, USA}
\email{mnajafi@mtu.edu}

\author{A.,Nayerhoda}
\affiliation{Institute of Nuclear Physics Polish Academy of Sciences, PL-31342 IFJ-PAN, Krakow, Poland}
\email{amid.nayerhoda@gmail.com}

\author{L.~Nellen}
\affiliation{Instituto de Ciencias Nucleares, Universidad Nacional Autónoma de Mexico, Ciudad de Mexico, Mexico}
\email{lukas@nucleares.unam.mx}


\author{N.~Omodei}
\affiliation{Department of Physics, Stanford University, Stanford, CA 94305–4060, USA}
\email{nicola.omodei@stanford.edu}

\author{M.,Osorio}
\affiliation{Instituto de Astronom\'{i}a, Universidad Nacional Autónoma de México, Ciudad de Mexico, Mexico }
\email{jmosorio@astro.unam.mx}

\author{E.~Ponce}
\affiliation{Facultad de Ciencias F\'{i}sico Matemáticas, Benemérita Universidad Autónoma de Puebla, Puebla, Mexico}
\email{eponce@fcfm.buap.mx}

\author{Y.~Pérez Araujo}
\affiliation{Instituto de F\'{i}sica, Universidad Nacional Autónoma de México, Ciudad de Mexico, Mexico}
\email{yuniorpy@gmail.com}

\author{E.G.~Pérez-Pérez}
\affiliation{Universidad Politecnica de Pachuca, Pachuca, Hgo, Mexico}
\email{egperezp@yahoo.com.mx}

\author[0000-0002-6524-9769]{C.D.~Rho}
\affiliation{Department of Physics, Sungkyunkwan University, Suwon 16419, South Korea}
\email{no397@naver.com}

\author{A.~Rodriguez Parra}
\affiliation{Universidad Michoacana de San Nicolás de Hidalgo, Morelia, Mexico}
\email{ancelmo.rodriguez@umich.mx}

\author{D.~Rosa-González}
\affiliation{Instituto Nacional de Astrof\'{i}sica, Óptica y Electrónica, Puebla, Mexico}
\email{danrosa@inaoep.mx}

\author{M.~Roth}
\affiliation{Los Alamos National Laboratory, Los Alamos, NM, USA}
\email{mattroth@lanl.gov}

\author{H.~Salazar}
\affiliation{Facultad de Ciencias F\'{i}sico Matemáticas, Benemérita Universidad Autónoma de Puebla, Puebla, Mexico}
\email{hsalazar@fcfm.buap.mx}


\author{A.~Sandoval}
\affiliation{Instituto de F\'{i}sica, Universidad Nacional Autónoma de México, Ciudad de Mexico, Mexico}
\email{asandoval@fisica.unam.mx}



\author{J.~Serna-Franco}
\affiliation{Instituto de F\'{i}sica, Universidad Nacional Autónoma de México, Ciudad de Mexico, Mexico}
\email{j_serna@ciencias.unam.mx}

\author{A.J.~Smith}
\affiliation{Department of Physics, University of Maryland, College Park, MD, USA}
\email{asmith8@umd.edu}

\author{Y.~Son}
\affiliation{University of Seoul, Seoul, South Korea}
\email{youngwan.son@cern.ch}

\author{R.W.~Springer}
\affiliation{Department of Physics and Astronomy, University of Utah, Salt Lake City, UT, USA}
\email{wayne.springer@utah.edu}

\author{O.~Tibolla}
\affiliation{Universidad Politecnica de Pachuca, Pachuca, Hgo, Mexico}
\email{omar.tibolla@gmail.com}

\author{K.~Tollefson}
\affiliation{Department of Physics and Astronomy, Michigan State University, East Lansing, MI, USA}
\email{tollefson@pa.msu.edu}

\author{I.~Torres}
\affiliation{Instituto Nacional de Astrof\'{i}sica, Óptica y Electrónica, Puebla, Mexico}
\email{ibrahim.torres23@gmail.com}

\author{R.~Torres-Escobedo}
\affiliation{Tsung-Dao Lee Institute, Shanghai Jiao Tong University, Shanghai, China}
\email{torresramiro350@sjtu.edu.cn}


\author{E.~Varela}
\affiliation{Facultad de Ciencias F\'{i}sico Matemáticas, Benemérita Universidad Autónoma de Puebla, Puebla, Mexico}
\email{enrique.varela@correo.buap.mx}

\author{L.~Villaseñor}
\affiliation{Facultad de Ciencias F\'{i}sico Matemáticas, Benemérita Universidad Autónoma de Puebla, Puebla, Mexico}
\email{lvillasen@gmail.com}

\author[0000-0001-6798-353X]{X.~Wang}
\affiliation{Department of Physics, Michigan Technological University, Houghton, MI, USA}
\email{xwang32@mtu.edu}

\author{Z.~Wang}
\affiliation{Department of Physics, University of Maryland, College Park, MD, USA}
\email[show]{zhen@umd.edu}

\author{I.J.~Watson}
\affiliation{University of Seoul, Seoul, South Korea}
\email{ian.james.watson@cern.ch}

\author{H.~Wu}
\affiliation{Department of Physics, University of Wisconsin-Madison, Madison, WI, USA}
\email{hwu298@wisc.edu}

\author{S.~Yu}
\affiliation{Department of Physics, Pennsylvania State University, University Park, PA, USA}
\email{sjy5345@psu.edu}

\author{H.~Zhou}
\affiliation{Tsung-Dao Lee Institute, Shanghai Jiao Tong University, Shanghai, China}
\email{hao_zhou@sjtu.edu.cn}

\author{C.~de León}
\affiliation{Universidad Michoacana de San Nicolás de Hidalgo, Morelia, Mexico}
\email{cederik.de.leon@umich.mx}

\collaboration{all}{HAWC Collaboration}

\begin{abstract}

Improving gamma-hadron separation is one of the most effective ways to enhance the performance of ground-based gamma-ray observatories. With over a decade of continuous operation, the High-Altitude Water Cherenkov (HAWC) Observatory has contributed significantly to high-energy astrophysics. To further leverage its rich dataset, we introduce a machine learning approach for gamma-hadron separation. A Multilayer Perceptron shows the best performance, surpassing traditional and other Machine Learning based methods. This approach shows a notable improvement in the detector's sensitivity, supported by results from both simulated and real HAWC data. In particular, it achieves a 19\% increase in significance for the Crab Nebula, commonly used as a benchmark. These improvements highlight the potential of machine learning to significantly enhance the performance of HAWC and provide a valuable reference for ground-based observatories, such as Large High Altitude Air Shower Observatory (LHAASO) and the upcoming Southern Wide-field Gamma-ray Observatory (SWGO).

\end{abstract}


\section{Introduction} \label{sec:intro} 

Recent breakthroughs in gamma-ray astronomy have greatly enhanced our understanding of high-energy astrophysical phenomena. Ground-based gamma-ray observatories, such as HAWC (\cite{albert2024performance}), LHAASO (\cite{bai2019large}) and High Energy Stereoscopic System (HESS) (\cite{ohm2023current}), play a crucial role in detecting and analyzing gamma-ray sources at Very-High Energy (50 GeV - 100 TeV) and Ultra-High Energy (100 TeV-100 PeV). When a gamma ray or cosmic ray enters the Earth's atmosphere, it initiates a shower of secondary particles, known as an Extensive Air Shower (EAS). A key challenge for ground-based observatories is the separation of gamma-ray-induced showers from the much larger ($>99.9\%$) background of cosmic-ray-induced hadronic showers.

The primary EAS components detectable by a ground-based gamma-ray observatory are electromagnetic particles, such as photons or electrons. Meanwhile, hadronic showers also produce a significant number of muons, which are typically generated from the decay of secondary charged pions and kaons. Besides their differing compositions, the distinct initial cascade mechanisms of gamma and hadronic showers also lead to differences in the spatial and energy distributions of their secondary particles. Based on these different shower characteristics, several well-established methods have been developed to distinguish gamma rays from the hadronic background. In recent years, Imaging Atmospheric Cherenkov Telescopes (IACTs) have successfully incorporated machine learning (ML) techniques to enhance gamma-hadron separation in practice (\cite{westerhoff1995separating, ohm2009gamma, albert2008implementation, kryukov2025machine}). Similar efforts have been reported by experiments like Astrophysical Radiation with Ground-based Observatory (ARGO) (\cite{pagliaro2011discrimination}) and LHAASO (\cite{wang2019gamma, li2025application}), although their studies were not tested on real data in their papers.

HAWC (\cite{alfaro2022gamma}) has also applied several ML techniques in gamma-hadron separation and tested them with real experimental data. In that study, the Boosted Decision Tree (BDT) algorithm demonstrated excellent performance, achieving approximately a 10\%  improvement in significance for sources such as the Crab. However, the performance of HAWC with BDT does not significantly surpass that of the recently optimized $\chi^{2}$ fit to the lateral distribution function in the HAWC Pass 5 reconstruction \citep{albert2024performance}. As a result, the BDT has not been adopted in the standard HAWC data analysis pipeline.

In this work, we apply a new ML-based approach to gamma-hadron separation in HAWC newly reconstructed Pass 5 data. Compared to the standard cut-based approach, the new ML-based gamma-hadron classification method demonstrates a significant improvement in sensitivity. Three major optimizations have been introduced in this study:

\begin{itemize} 
\item \textbf{Expanded feature set:} Previous work utilized only seven input features, whereas this study incorporates 20 features, capturing more detailed event-level information relevant to classification. The details about the features are described in Section~\ref{subsec:input}. 
\item \textbf{Unified training strategy:} Instead of training separate models for low-, medium-, and high-energy events as done in earlier studies, we train a single model using the gamma ray simulation set  across all energy ranges from 5 GeV to 500 TeV with cosmic ray simulation set as background. The details are described in Section \ref{sec:data}. 
\item \textbf{More complex ML model:} This study employs complex architectures with more hidden layers than those used in previous approaches. Models are described in Section~\ref{sec:methods}.
\end{itemize}

In this paper, Section~\ref{sec:hawc} provides a brief introduction to the HAWC Observatory. Sections~\ref{sec:data} and~\ref{sec:methods} describe the datasets and machine learning models used for training and evaluation. Section~\ref{sec:perf} presents a performance comparison between the standard cut and ML methods for gamma-hadron separation. Finally, Section~\ref{sec:conc} provides the conclusions of this study.

\section{HAWC observatory} \label{sec:hawc}

The High-Altitude Water Cherenkov (HAWC) Observatory is a ground-based gamma-ray detector located at an altitude of 4,100 meters above sea level, at coordinates $18.99^\circ{N},\ 97.31^\circ{W}$, in Mexico (\cite{abeysekara2023high}). The primary array consists of 300 Water Cherenkov Detectors (WCDs). The size of each WCD is 5.4 meters high and 7.3 meters in diameter, deployed over a total physical area of approximately 22,000 m². For the past decade, HAWC has been continuously observing the northern sky, detecting EAS produced by high-energy cosmic rays and gamma rays. The sensitivity of HAWC spans from hundreds of GeV to over 100 TeV, making it an ideal tool for studying high-energy astrophysical sources. Previous studies have demonstrated its effectiveness in identifying gamma-ray sources, detecting transient events, and searching for new physics. The recent HAWC Pass 5 reconstruction has significantly enhanced the performance of the observatory, as described in (\cite{albert2024performance}). Building on this upgraded framework, the following studies will be conducted using data and simulation processed with the improved Pass 5 reconstruction. 

\section{Data sets}\label{sec:data}

Two types of datasets are used in this work: Monte Carlo (MC) data and real HAWC data, as described in Sections~\ref{subsec:mcdata} and~\ref{subsec:realdata}, respectively. During training, both MC gamma-ray and hadronic events are used. For evaluation, MC gamma-ray events are used as the signal, while real HAWC data serves as the background.

\subsection{Monte Carlo data}\label{subsec:mcdata}

Simulated EAS were generated using CORSIKA (\cite{heck1998corsika}), incorporating the QGSJET-II-04 and FLUKA hadronic interaction models. The simulations included nine particle species: gamma rays and cosmic rays (protons, helium, carbon, oxygen, neon, magnesium, silicon, and iron nuclei). The gamma-ray energies ranged from 5 GeV to 500 TeV, while the cosmic-ray energies spanned from 5 GeV to 2 PeV. All particles were distributed isotropically across the sky with zenith angles below $60^{\circ}$. These events were then processed using HAWC simulation software based on GEANT4 (\cite{allison2007facilities}), to propagate the ground-level particles through the HAWC tanks.

The simulated trigger events were subsequently reconstructed using the same offline pipeline applied to real HAWC data. As a result, the MC dataset adopts the same data structure and format as real observations, while retaining the true shower information. Finally, the reconstructed events were reweighted according to their particle species, spectral distributions, and shower core weights assigned during the simulation process. Further details can be found in (\cite{abeysekara2023high}).

\subsection{Real HAWC data}\label{subsec:realdata}

We selected a 3-year dataset as the background sample for evaluating the gamma-hadron separation performance and sensitivity curves, as described in Section~\ref{subsection:Qfactor} and Section~\ref{subsection:sensitivity}. Events in this dataset are filtered based on the fraction of hit channels (fHit). Partial events with low fHit values are excluded. This selection ensures that at least hundreds of high-energy events are retained after gamma-hadron separation cut, while keeping the dataset small enough to be handled conveniently. Each event will be assigned a weight to preserve the original distribution of fHit. 

In Section~\ref{sec:perf}, a dataset comprising approximately 3070 days of observations from 2013 to 2024 is used to compare performance for several representative astrophysical sources. In this study, we present detailed results for the Crab Nebula, which serves as a standard reference for all TeV gamma-ray instruments. To evaluate the declination dependence of gamma-hadron separation, we also compare the detection significances of the Galactic Center and Boomerang. The Galactic Center lies near the southern edge of HAWC field of view (FOV) with a zenith angle of $\approx 48^{\circ}$ (\cite{albert2024observation}), while Boomerang (\cite{alfaro2024testing}) is a prominent bright source located near the northern edge with a zenith angle of $\approx 42^{\circ}$. 

\section{Model Construction}\label{sec:methods}
\subsection{Input features} \label{subsec:input} 

In addition to the analysis method we used in Pass 5 (\cite{albert2024performance}), we employ three ML-based models to discriminate gamma rays from the hadronic background. All three methods share the same training and evaluating procedures; the only difference lies in the architecture of the ML models.

Each ML-based method uses 20 input features, which are categorized as follows:

\begin{itemize} 
\item \textbf{R, Zenith:} 

\textit{$R$} represents the distance from the shower core to the center of the detector array. 

\textit{Zenith} denotes the zenith angle of the incoming shower.

\item \textbf{fHit, fTank, NNEnergy, GPEnergy:} These features relate to the energy of the shower.  

\textit{fHit} is fraction of hit channels, defined as $fHit = nHit / nCh$, where $nHit$ is the triggered number of PMTs, and $nCh$ is the total number of operational PMTs. \textit{fHit} serves as a good proxy for energy at the lower end of our energy range. However, due to detector saturation, its reliability diminishes above tens of TeV. To provide accurate energy estimation and extend energy range, we adopt a 2D approach in analysis, combining fHit with \textit{NNEnergy} or \textit{GPEnergy} in analysis (\cite{abeysekara2019measurement}). 

\textit{fTank} is fraction of hit tanks, defined as $fTank = nTank / nTankCh$, where $nTank$ is the triggered number of tanks, and $nTankCh$ is the total number of operational tanks.  

\textit{NNEnergy} and \textit{GPEnergy} are two energy estimators described in (\cite{abeysekara2019measurement}). The GP algorithm estimates energy based primarily on the charge density at an optimal distance from the shower axis, while the NN method employs a neural network to infer the primary energy.

\item \textbf{LDF-$\chi^{2}$, LDFAmp:}  

\textit{LDF-$\chi^{2}$} is the reduced chi-square from fitting a lateral distribution function (LDF) derived from a modified version of the Nishimura–Kamata–Greisen (NKG) function~(\cite{kamata1958lateral}), in which a factor of $1/r$ is introduced to steepen the distribution. This modification reflects the fact that the original NKG function describes particle density, while HAWC is sensitive to energy density, which falls off more steeply by a factor of $1/r$ from the shower core.

\textit{LDFAmp} is the logarithm of the fitted LDF amplitude. 

\item \textbf{PINCness, Compactness:}  

\textit{PINCness} quantifies the smoothness of the lateral charge distribution(\cite{abeysekara2017observation}).  

\textit{Compactness} is defined as the ratio of the number of PMTs hit in an EAS event to the maximum charge recorded by any PMT located more than 40 meters from the shower core (\cite{atkins2003observation}).

\item \textbf{fAnnulusQ0 to fAnnulusQ9:}  
These ten features represent the normalized charge distribution measured in concentric annular regions centered on the shower core. The innermost region (fAnnulusQ0) is a circle with a radius of 10 meters. Each subsequent region (fAnnulusQ1 to fAnnulusQ9) is an annulus with a width of 10 meters, extending outward from the edge of the previous region. The total charge across all ten regions is normalized such that the sum of these features equals 1.

\end{itemize}

\subsection{Standard Cut (SC)}\label{sub:SC}     
We hereafter refer to the current gamma-hadron separation method used in Pass 5 reconstruction as the Standard Cut (SC). The SC method uses a specific combination of features, including LDF-$\chi^{2}$, PINCness, Compactness, and Zenith angle in different fHit bins. Further details are provided in (\cite{albert2024performance}).

\subsection{Boosted Decision Trees (BDT)} \label{sub:bdt} 

   Decision trees are a simple yet powerful type of model that classify events by applying a series of binary decisions, ultimately assigning them to leaf nodes. They are commonly used in high-energy physics for tasks such as distinguishing signal events (e.g., gamma rays) from background (e.g., hadrons).

 In this work, we implement a Gradient Boosting algorithm for our BDT model using the Toolkit for MultiVariate data Analysis (TMVA) (\cite{hocker2008tmva}) in the ROOT framework (\cite{root}). The term 'Boosted' refers to the technique of combining many weak decision trees, where each new tree focuses on correcting the mistakes of the previous ones. This sequential improvement process leads to a much stronger model. Gradient boosting focuses on misclassified events by adjusting weights iteratively, improving performance with each new tree. To achieve good performance while avoiding overfitting, we select the following parameters based on initial parameter scans and empirical validation:
 
\begin{itemize}
    \item \textbf{Number of trees = 500:} A larger number of trees improves the model's ability to capture complex patterns. We limit it to 500 to balance accuracy and training time.
    \item \textbf{Maximum tree depth = 6:} A depth of 6 allows the model to make moderately complex decisions without overfitting, based on our validation tests.
    \item \textbf{Learning rate  = 0.2:} This controls how much each new tree corrects the previous ones. A smaller rate ensures stable learning, and 0.2 was found to be an effective compromise.
    \item \textbf{Bagged boosting with 50\% sample fraction:} Each tree is trained on a random subset (50\%) of the data to reduce overfitting and improve generalization.
    \item \textbf{Number of node-splitting cuts = 20:} This controls how finely the algorithm searches for split points at each node. We chose 20 to match the number of input variables, providing a balance between resolution and simplicity.
\end{itemize}

 \subsection{Multilayer Perceptron (MLP)} \label{sub:mlp}

Neural networks are powerful tools for classification and have become increasingly common in high-energy physics. In this work, we implement a Multilayer Perceptron (MLP) using the PyTorch software package (\cite{paszke2019pytorch}) to distinguish between gamma-ray and hadron events. A MLP is a type of feedforward neural network composed of multiple layers of interconnected nodes (neurons). Each layer processes the data and passes information to the next, enabling the model to learn complex patterns in the input features. PyTorch is a popular open-source tool that makes it easy and efficient to build and train neural networks.

Our MLP model consists of the following architectures:

\begin{itemize}
    \item \textbf{Input layer:} Accepts 20 input features to match the number of variables provided by each event.
    \item \textbf{First and second hidden layers:} Consist of 128 and 64 neurons, respectively. These sizes were chosen empirically: 128 neurons allow the model to learn complex interactions between features, while 64 neurons in the second layer help refine these representations without excessive complexity.
    \item \textbf{Dropout layers:} Applied after each hidden layer with a dropout rate of 20\%. This technique randomly disables 20\% of neurons during training, helping to reduce overfitting and improve generalization.
    \item \textbf{Output layer:} A single neuron with sigmoid activation, which outputs a probability score for binary classification between gamma and hadron events.
    \item     \textbf{Learning rate  = 0.01:} This a value chosen after testing several options and finding it provided a good balance between convergence speed and training stability. 
\end{itemize}

    We train the model using the Binary Cross-Entropy loss (BCELoss) (\cite{paszke2019pytorch}) and the Adam optimizer (\cite{kingma2014adam}). The BCELoss evaluates how well the predicted probabilities match the true binary labels, with lower loss values indicating better model performance.  Adam optimizer is an optimization algorithm that updates the model’s weights by combining momentum and adaptive learning rate techniques, providing efficient and stable training.

 \subsection{Convolutional Neural Networks (CNN)}\label{sub:cnn}

Convolutional Neural Networks (CNN) are widely used in image and pattern recognition tasks due to their ability to automatically identify spatial structures in data. A CNN model can recognize important features such as shapes or patterns.  In this work, we use PyTorch to develop a hybrid model combining CNN and MLP to classify gamma-ray and hadron events by  leveraging the event characteristics, particularly the spatial structure of the event shower.

The model consists of two parallel branches:

\begin{itemize}
    \item \textbf{CNN branch:} Processes the annulus charge distribution features using two convolutional layers, each followed by a max-pooling operation to reduce dimensionality while preserving important features. After the convolutional layers, the resulting feature maps are flattened to prepare for integration with the MLP branch.
    \item  \textbf{MLP branch:} Processes the remaining non-annulus features through two fully connected layers. These layers help extract interactions between features that are not spatially organized.
    \item \textbf{Fusion and classification:} The outputs of both branches are concatenated and passed through a final dense layer with a sigmoid activation function, which produces a probability score for classifying the event as gamma or hadron.
\end{itemize}

All other settings, including dropout rates and activation functions, are consistent with the MLP model described previously.

Although CNN are typically used for 2D image data, in our case we apply them to a 1D vector of features (fAnnulusQ0 to fAnnulusQ9). This vector represents the radial energy distribution of the air shower, which exhibits spatial patterns similar to those in images. By using 1D convolutions, the CNN can learn and extract relevant local features from this structured input. This represents an initial attempt to apply CNN to this type of data in the context of gamma-hadron classification.

\subsection{Model assessment with Q-factor}
\label{subsection:Qfactor}
To evaluate the performance of the gamma-hadron separation models, we defined a quality factor (Q-factor), as shown in Equation~\ref{equ:qfact} below:
\begin{equation}
\label{equ:qfact} 
Q = \frac{\xi_{\gamma}}{\sqrt{\xi_{h}}},
\end{equation}

Where $\xi_{\gamma}$ represents the gamma efficiency, which is the percentage of gamma-ray showers that pass the gamma-hadron separation criteria. Similarly, $\xi_{h}$ denotes the hadron efficiency, referring to the percentage of hadron showers that pass the same criteria. This equation assumes a sufficiently large background, which is validated in this study. 

In ground-based gamma-ray observatories, Q-factor is commonly used as a metric to assess the effectiveness of event selection criteria in enhancing the detection significance of astrophysical sources. By optimizing the Q-factor, one can identify the most effective separation between gamma-ray signals and hadronic background.

The ML-based models provide a probability of being a gamma-like event ($p_{\gamma}$ hereafter) for each shower. The optimal Q-factor search involves scanning the $p_{\gamma}$ value from 0 to 1, where 0 indicates the least likelihood of being a gamma-ray shower, and 1 indicates the highest likelihood. To simplify the comparison, we set the gamma efficiency value in SC model as the lower limit, and then search the best Q-factor by scanning the $p_{\gamma}$ cut.

\begin{figure}[ht!] 
    \centering
    \subfloat{%
        \includegraphics[width=0.48\textwidth]{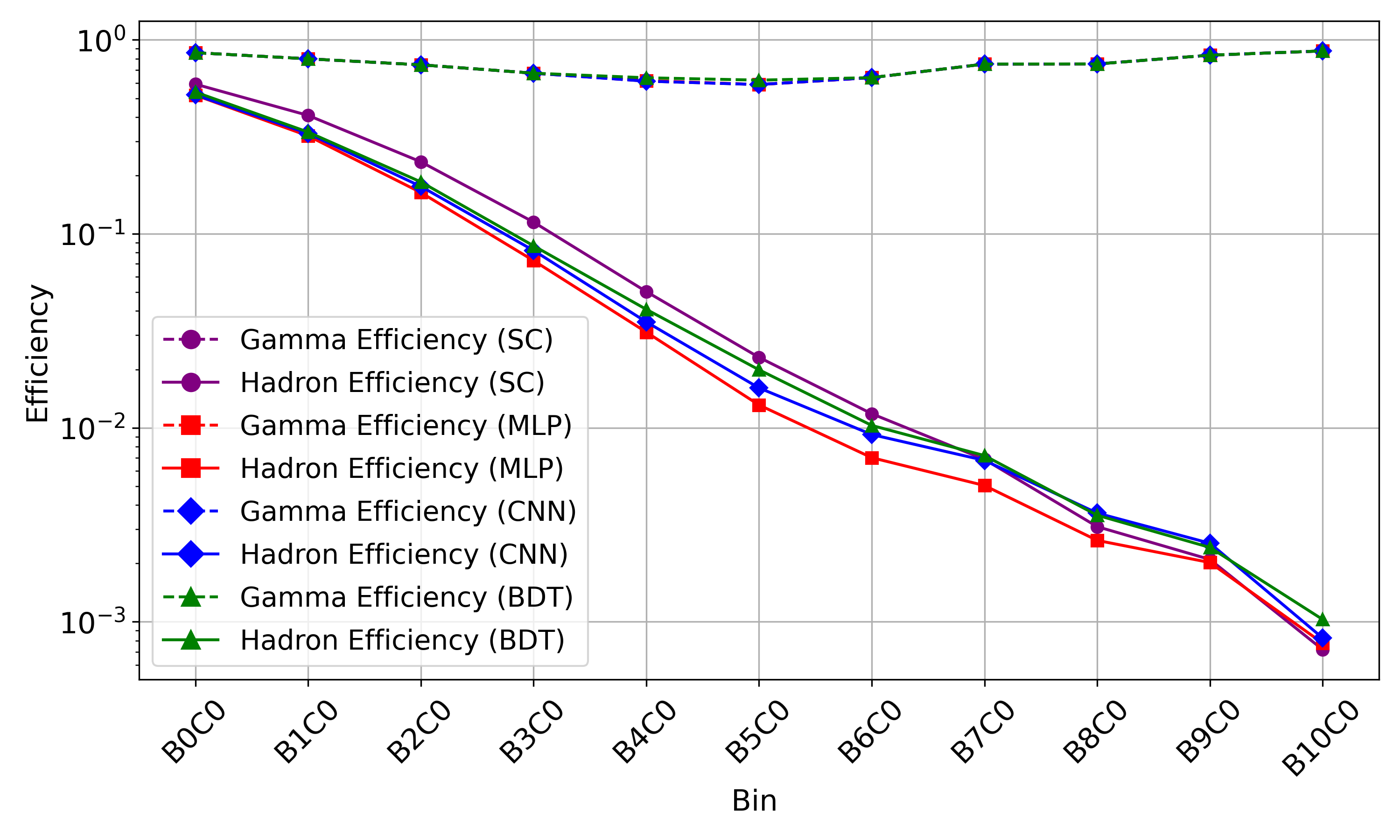}%
        \label{fig:eff_c0}%
    }%
    \hfill%
    \subfloat{%
        \includegraphics[width=0.48\textwidth]{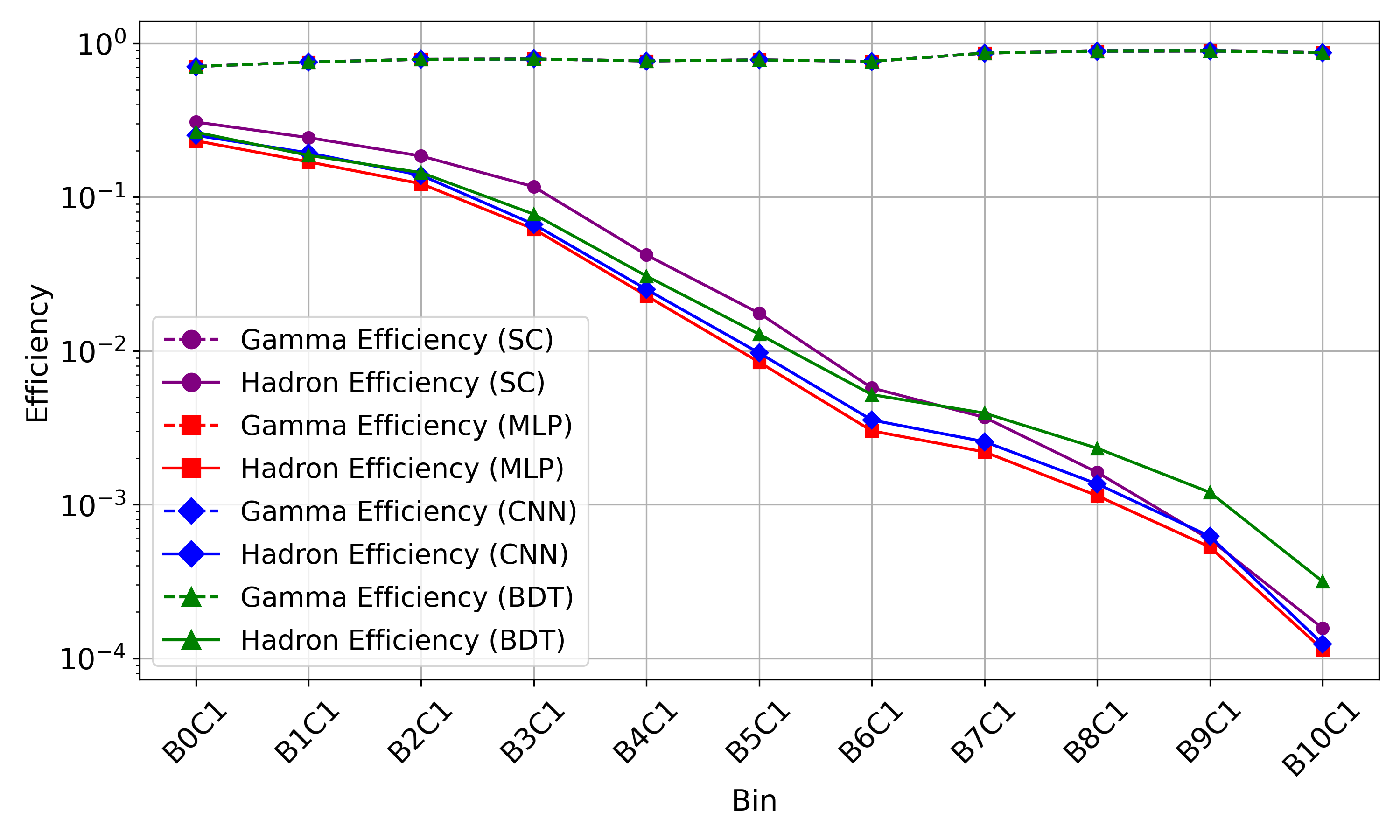}%
        \label{fig:eff_c1}%
    }%
    \caption{Gamma-ray efficiencies (dashed lines) and hadron efficiencies (solid lines) as a function of fHit bins for different classification methods. (a): Results for on-array events. (b): Results for off-array events. Classification methods compared include SC, MLP, CNN, and BDT.}
    \label{fig:eff}
\end{figure}

\begin{figure}[ht!] 
    \centering
    \subfloat{%
        \includegraphics[width=0.48\textwidth]{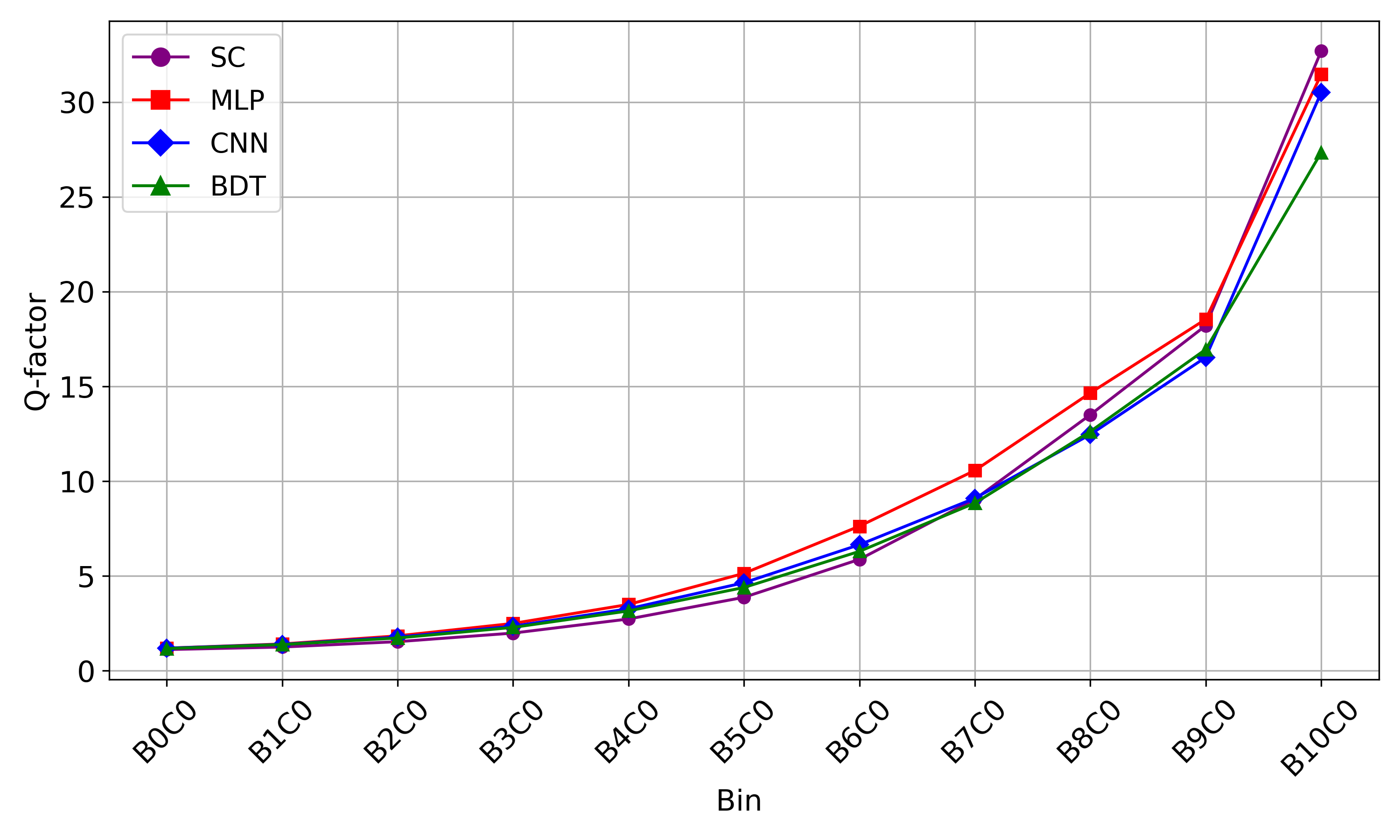}%
        \label{fig:eff_qf_c0}%
    }%
    \hfill%
    \subfloat{%
        \includegraphics[width=0.48\textwidth]{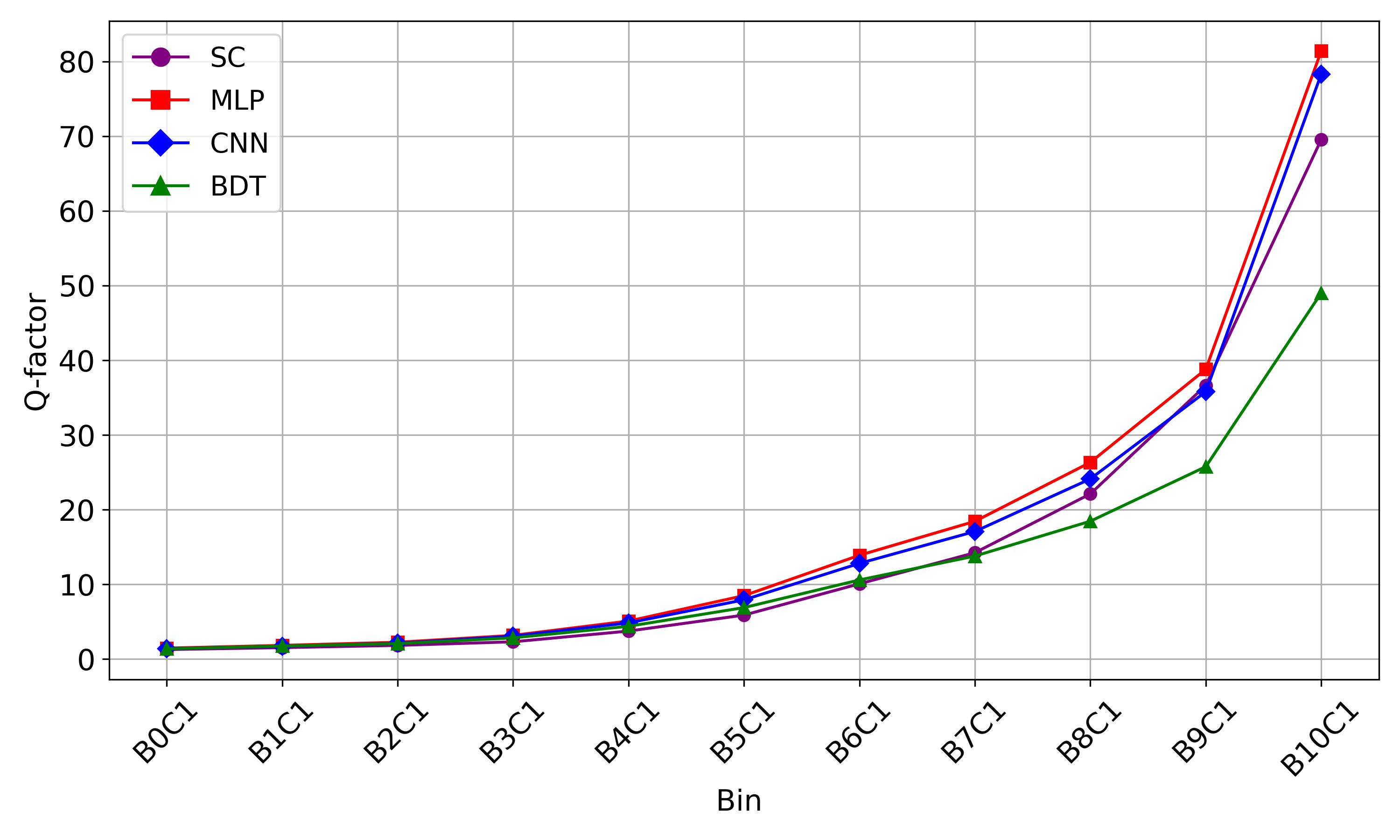}%
        \label{fig:eff_qf_c1}%
    }%
    \caption{Q-factor as a function of fHit bins for different classification methods. (a): Results for on-array events. (b): Results for off-array events. Classification methods compared include SC, MLP, CNN, and BDT. Q-factors improve with increasing fHit bins, with notable differences across methods, especially in high-fHit bins.}
    \label{fig:qfactor}
\end{figure}

The results are presented in Figure~\ref{fig:eff} and Figure~\ref{fig:qfactor}. Shower events are divided into two groups: on-array events and off-array events. On-array events are defined as those whose shower cores are located in the main array area (about 100-meter radius of the main array center). Off-array events refer to events whose shower cores fall outside the on-array region but still within 150 meters of the array center. Figure~\ref{fig:eff} shows the gamma-ray (dashed lines) and hadron (solid lines) efficiencies as a function of fHit. Figure~\ref{fig:qfactor} displays the corresponding Q-factors across different fHit bins. As shown in Figures~\ref{fig:eff_qf_c0} and~\ref{fig:eff_qf_c1}, higher fHit bins—corresponding to higher-energy events—generally yield better Q-factors. All four classification methods described in Section~\ref{sec:methods} are compared. Among them, the MLP model achieves the best Q-factors in nearly all energy bins, indicating superior hadron rejection performance (\cite{abeysekara2019measurement}).

The performance of the MLP model is evaluated using the Area Under the Curve (AUC) of the Receiver Operating Characteristic (ROC) \citep{fawcett2006introduction} as the primary metric. ROC curves illustrate the trade-off between the true positive rate (TPR, or gamma efficiency here) and false positive rate (FPR, or hadron efficiency here) at different classification thresholds, helping to assess how well the model distinguishes gamma-ray from hadron events. The AUC summarizes this performance with a single value, where 1 indicates perfect classification and 0.5 corresponds to random guessing. A higher AUC therefore reflects better separation ability.

\begin{figure}[ht!] 
    \centering
    \subfloat{%
        \includegraphics[width=0.48\textwidth]{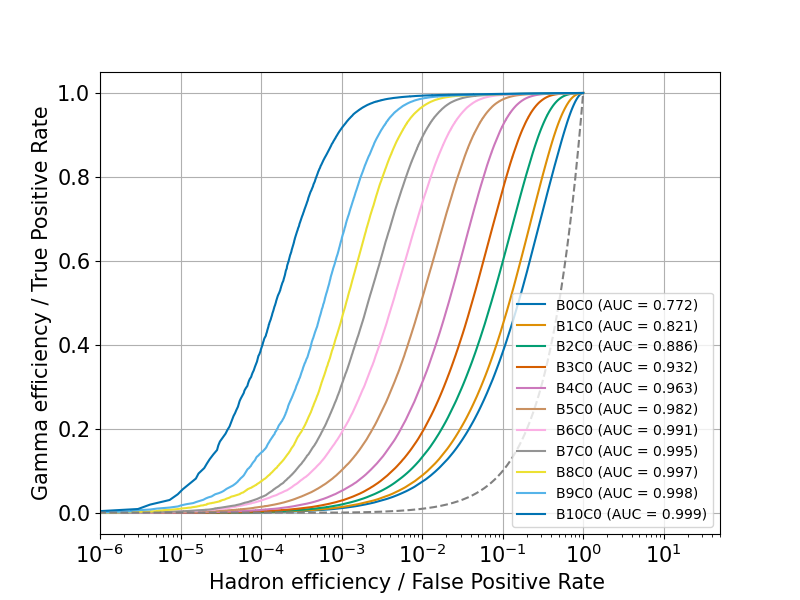}%
        \label{roc_c0}%
    }%
    \hfill%
    \subfloat{%
        \includegraphics[width=0.48\textwidth]{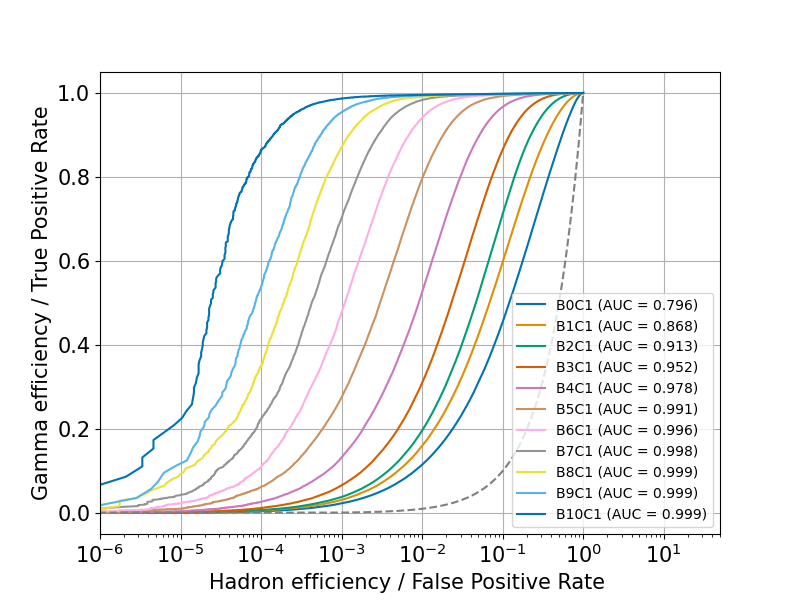}%
        \label{roc_c1}%
    }%
    \caption{Receiver Operating Characteristic (ROC) curves and Area Under the Curve (AUC) values for MLP model performance across different fHit bins. Left panel (a): Results for on-array events. Right panel (b): Results for off-array events. Each curve corresponds to a specific fHit bin, labeled from B0 to B10. The MLP model exhibits progressively better classification performance with increasing fHit, as indicated by higher AUC values.}
    \label{fig:ROC_curve}
\end{figure}

Figure~\ref{fig:ROC_curve} presents the ROC curves of the performance
of the MLP model, with Figure~\ref{roc_c0} showing results for on-array events and Figure~\ref{roc_c1} for off-array events. All fHit bins from 0 to 10 are included, and their corresponding median energies are listed in Table~\ref{tab:crab_significance_comparison}. A solid ROC curve lying below the gray dashed line indicates performance worse than random guessing (i.e., classification accuracy below 50\%). As shown in Figure~\ref{fig:ROC_curve}, all curves lie above the 50\% accuracy threshold, indicating meaningful classification performance. As expected, higher-energy events demonstrate better gamma-hadron separation, reflected in higher ROC curves.

Notably, off-array events show better overall performance for the MLP model relative to the SC method compared to on-array events, further demonstrating the robustness of the MLP model across different event categories. That is, although off-array events are harder to classify due to their peripheral geometry and reduced signal quality, the model still delivers a notable performance improvement for them. This indicates that the MLP model is not overfitting to the dominant (on-array) training data. The MLP architecture is capable of learning event-level features, such as the spatial and temporal distributions of secondary particles, which contributes to improved classification accuracy. The trained model is subsequently applied to experimental data, replacing the traditional SC method.

\section{Data Performance} \label{sec:perf}

\subsection{Sensitivity of Crab-like point source}
\label{subsection:sensitivity}
 
\begin{figure}[ht!]
\begin{centering}
\includegraphics[width=0.7\textwidth]{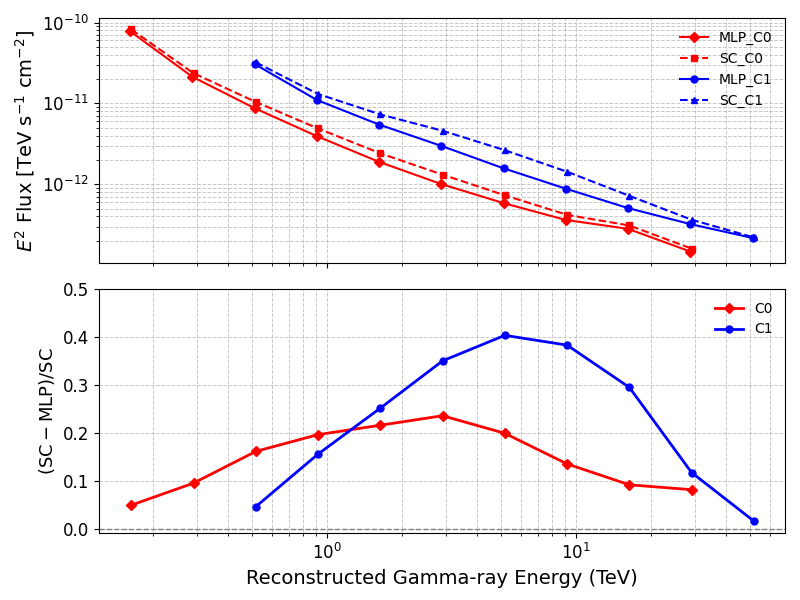}
\caption{HAWC 10-year sensitivity comparison between MLP and SC methods. Top: Comparison of differential sensitivity curves between the MLP-based event selection (solid lines) and the SC method (dashed lines) as a function of reconstructed gamma-ray energy. Bottom: Relative improvement in sensitivity, calculated as $(Flux_{\rm sc}-Flux_{\rm  mlp})/Flux_{\rm sc}$.}
\label{fig:sens_nn}
\end{centering}
\end{figure}

We applied the MLP method to real HAWC data collected between 2013 and 2024, covering approximately 3,070 transits. The resulting sensitivity of the HAWC Observatory shows improvements of about 23\%  (on-array events) and 40\% (off-array events) compared to the result with SC method for gamma-hadron separation separately. Figure~\ref{fig:sens_nn} presents a comparison of the sensitivity curves across all fHit energy bins, clearly demonstrating the gains achieved through ML–based event classification. The MLP-based sensitivity consistently surpasses that of the SC method, particularly across the entire energy range, especially in the medium energy bins. This advantage is crucial for detecting weak gamma-ray sources.

The MLP approach demonstrates a clear advantage at medium energies, particularly in the 1--20 TeV range, where it achieves notable improvements over the SC method. This improvement arises from the ability of machine learning models to simultaneously incorporate multiple variables in the gamma-hadron separation process. However, at both the low and high ends of the energy spectrum, the performance gap narrows. Below 1~TeV, the intrinsic similarities between gamma-ray and hadron-induced air showers make the separation task inherently difficult due to the underlying physics of shower development. At energies above tens of TeV, the performance of the MLP and SC methods becomes comparable. This is primarily because the muon content in hadronic showers increases with energy, serving as an especially effective discriminator between gamma rays and hadrons (\cite{tian2018study,aharonian2021observation}). Furthermore, the SC method includes explicit optimization with respect to declination at high energies, while the current MLP implementation does not incorporate such directional dependence.  We also examined declination dependence in Section~\ref{subsection:GC_Boom}.

\subsection{Crab}
\label{subsection:crab}
We assessed the significance and spectral energy distribution (SED) of the Crab using both the SC and MLP methods. This analysis was conducted with the HAWC dataset described in Section~\ref{subsection:sensitivity}. 
The MLP method produces a peak significance of $356\sigma$, in contrast to a lower peak significance of $299\sigma$ from the SC method
, reflecting a 19\% enhancement, consistent with expectations based on sensitivity curve comparisons. These findings confirm that the application of machine learning techniques boosts the statistical confidence of gamma-ray detection.

Table~\ref{tab:crab_significance_comparison} presents a bin-by-bin comparison of the detection significance for the Crab using the MLP and the SC approaches. The table includes both on-array and off-array fHit bins, along with their corresponding median energies. The MLP method demonstrates clear improvements in most bins, particularly in the mid- to high-energy range. The most significant enhancement is observed in bin B5C1, with a 53\% increase in significance compared to SC. While the lowest-energy bin (B0C0) shows a slight decrease (which bin is often excluded from analyses due to its poor performance), the majority of bins exhibit consistent gains ranging from a few percent up to over 50\%. These results highlight the effectiveness of the MLP approach in enhancing the sensitivity of HAWC, especially for medium energy and off-array events. Furthermore, the observed improvements align well with the sensitivity enhancements discussed in Section~\ref{subsection:sensitivity}.


\begin{table}[ht!]
    \centering
    \begin{tabular}{ccccc}
        \toprule
        \textbf{fHit Bins}  & \textbf{Median Energy(TeV)} & \textbf{$\sigma_{\rm mlp}$}& \textbf{$\sigma_{\rm sc}$}& \textbf{($\sigma_{\rm mlp}$ - $\sigma_{\rm sc}$)/$\sigma_{\rm sc}$}\\
        \midrule
        B0C0 &  0.28  &  6.60  & 6.65  & -0.75\% \\
        B1C0 & 0.38  & 21.75  & 19.57 & 11.14\% \\
        B2C0 & 0.53 & 39.03  & 34.48 & 13.20\% \\
        B3C0 & 0.83 & 71.21  & 57.69 & 23.45\% \\
        B4C0 & 1.37 & 113.84 & 88.15 & 29.15\% \\
        B5C0 & 2.25 & 139.99 & 110.66 & 26.50\% \\
        B6C0 & 3.68 & 142.99 & 120.50 & 18.71\% \\
        B7C0 & 5.97 & 122.23 & 119.46 & 2.32\% \\
        B8C0 & 9.54 & 102.95 & 94.66 & 8.74\% \\
        B9C0 & 14.63 & 79.41  & 74.69 & 6.32\% \\
        B10C0 & 30.46 & 77.91 & 74.28 & 4.89\% \\
        \midrule
        B0C1 & 0.57  & 12.13  & 12.00 & 1.08\% \\
        B1C1 & 0.88 & 21.69  & 18.27 & 18.75\% \\
        B2C1 & 1.29 & 35.89  & 29.64 & 21.09\% \\
        B3C1 & 2.02 & 52.15  & 39.67 & 31.39\% \\
        B4C1 & 3.66 & 68.72  & 47.24 & 45.48\% \\
        B5C1 & 6.21 & 80.28  & 52.15 & 53.99\% \\
        B6C1 & 10.27 & 79.69  & 57.54 & 38.60\% \\
        B7C1 & 16.62 & 60.49  & 50.69 & 19.29\% \\
        B8C1 & 25.78 & 46.49  & 41.98 & 10.73\% \\
        B9C1 & 41.47 & 29.82  & 29.16 & 2.26\% \\
        B10C1 & 73.91 & 32.01  & 30.43 & 5.19\% \\

        \midrule
        \bottomrule
    \end{tabular}
    \caption{Comparison of detection significances for the Crab using the MLP ($\sigma_{\mathrm{MLP}}$) and the SC ($\sigma_{\mathrm{SC}}$) across different fHit bins. The table also shows the relative improvement $(\sigma_{\mathrm{MLP}} - \sigma_{\mathrm{SC}})/\sigma_{\mathrm{SC}}$ for both on-array (C0) and off-array (C1) events. The median energy of each bin is also listed. HAWC Energy bin defined in (\cite{abeysekara2019measurement})}
    \label{tab:crab_significance_comparison}
\end{table}

\begin{figure}[ht!]
\begin{centering}
\includegraphics[width=0.7\textwidth]{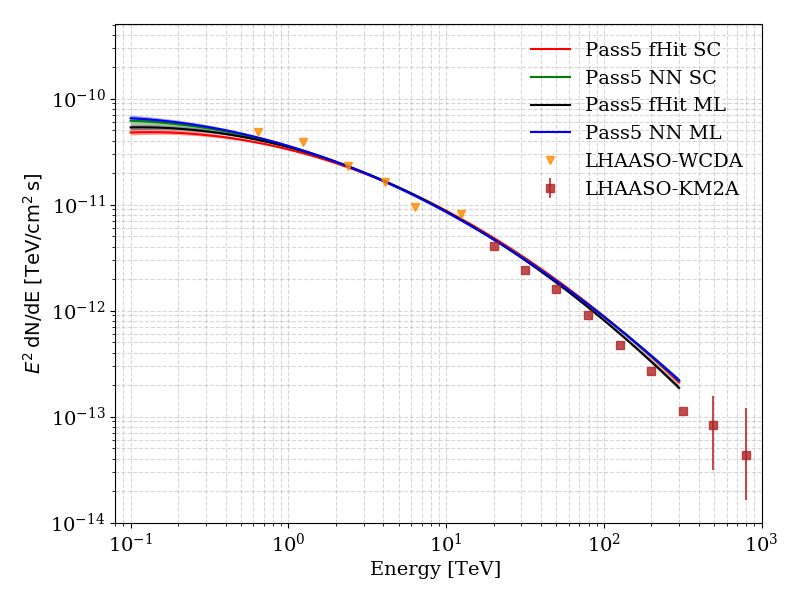}
\caption{Spectral energy distribution of the Crab Nebula. The plot compares different analysis methods and datasets, including Pass 5 fHit bins with ML (black line), Pass 5 fHit bins with SC (red line), Pass 5 NN with SC (green line) and Pass 5 NN ML (blue line). The HAWC Pass 5 NN SC and NN ML maps were produced using a 2D approach that combines fHit and NNEnergy. The details of HAWC energy estimatiors are described in (\cite{abeysekara2019measurement}). LHAASO-WCDA and LHAASO-KM2A data are represented by orange triangles and brown squares, respectively (\cite{lhaaso2021peta}).}
\label{fig:crab_spectrum}
\end{centering}
\end{figure}

Figure~\ref{fig:crab_spectrum} shows the SED of the Crab, emphasizing the consistency achieved through the adoption of MLP in gamma-hadron separation. The flux measured using the MLP (black line and blue line) closely match that from the SC method (red line and green line), demonstrating the reliability of the ML approach. 

\subsection{Galactic Center and Boomerang} 
\label{subsection:GC_Boom}

To assess the declination dependence on gamma-hadron separation performance, we selected the Galactic Center and Boomerang as representative sources. The Galactic Center lies near the southern edge of FOV of HAWC, while Boomerang  is a prominent, bright source located near the northern edge.

The Galactic Center, although relatively difficult to observe due to its low elevation in HAWC's sky coverage, shows a notable improvement when analyzed with the MLP method. The detection significance reaches 8.64$\sigma$ with MLP, compared to 6.43$\sigma$ with SC method (\cite{albert2024observation}), corresponding to a 34.3\% enhancement.

Boomerang, a consistently bright and high-declination source, also benefits from the MLP-based analysis, though to a lesser extent. The detection significance increases from 14.40$\sigma$ (\cite{alfaro2024testing}) with SC to 15.17$\sigma$ with MLP, yielding a 5.3\% improvement.

\section{Conclusion and Discussion} \label{sec:conc}

In this study, we successfully implemented BDT, MLP and CNN models for gamma-hadron separation using data from HAWC. Our results demonstrate a significant enhancement in sensitivity, particularly through the use of a MLP model, which consistently outperformed the traditional SC method. The improved performance is evident in both simulated and real HAWC data, with notable increases in source detection significance. Specifically, with the HAWC data of 3,070 days, the MLP model achieved an approximate 19\% increase in significance for the benchmark source, the Crab, compared to the SC method.

The consistency of Crab SED analysis, as shown in Figure \ref{fig:crab_spectrum}, between the ML and SC methods
highlights the robustness and reliability of the ML approach in preserving the spectral characteristics of high-energy astrophysical sources.

Our study emphasizes the potential of ML techniques to advance the observational capabilities of ground-based gamma-ray observatories. The ability of the MLP model to enhance sensitivity, especially in the low-energy regime, is critical for the detection of weak gamma-ray sources. This improvement is attributed to the model's capability to leverage complex event-level features, thereby improving the separation of gamma-ray signals from hadronic background noise.

Future work will focus on refining the ML model by expanding the analysis to the HAWC dataset including outrigger data which has an extended energy range (\cite{marandon2019latest}). This will further explore the applicability of ML techniques in gamma-ray astronomy, potentially paving the way for future advancements in the field.

In conclusion, the integration of machine learning into HAWC data analysis significantly enhances the observatory’s sensitivity and sets a strong precedent for the adoption of similar techniques in ground-based particle sampling experiments, such as LHAASO and the upcoming SWGO.

\begin{acknowledgments}
We acknowledge the support from: the US National Science Foundation (NSF); the US Department of Energy Office of High-Energy Physics; the Laboratory Directed Research and Development (LDRD) program of Los Alamos National Laboratory; Consejo Nacional de Ciencia y Tecnolog\'{i}a (CONACyT), M\'{e}xico, grants LNC-2023-117, 271051, 232656, 260378, 179588, 254964, 258865, 243290, 132197, A1-S-46288, A1-S-22784, CF-2023-I-645, c\'{a}tedras 873, 1563, 341, 323, Red HAWC, M\'{e}xico; DGAPA-UNAM grants IG101323, IN111716-3, IN111419, IA102019, IN106521, IN114924, IN110521 , IN102223; VIEP-BUAP; PIFI 2012, 2013, PROFOCIE 2014, 2015; the University of Wisconsin Alumni Research Foundation; the Institute of Geophysics, Planetary Physics, and Signatures at Los Alamos National Laboratory; Polish Science Centre grant, 2024/53/B/ST9/02671; Coordinaci\'{o}n de la Investigaci\'{o}n Cient\'{i}fica de la Universidad Michoacana; Royal Society - Newton Advanced Fellowship 180385; Gobierno de España and European Union NextGenerationEU, grant CNS2023- 144099; The Program Management Unit for Human Resources \& Institutional Development, Research and Innovation, NXPO (grant number B16F630069); Coordinaci\'{o}n General Acad\'{e}mica e Innovaci\'{o}n (CGAI-UdeG), PRODEP-SEP UDG-CA-499; Institute of Cosmic Ray Research (ICRR), University of Tokyo. H.F. acknowledges support by NASA under award number 80GSFC21M0002. C.R. acknowledges support from National Research Foundation of Korea (RS-2023-00280210). We also acknowledge the significant contributions over many years of Stefan Westerhoff, Gaurang Yodh and Arnulfo Zepeda Dom\'inguez, all deceased members of the HAWC collaboration. Thanks to Scott Delay, Luciano D\'{i}az and Eduardo Murrieta for technical support.
\end{acknowledgments}




\bibliography{sample7.bib}

\begin{thebibliography}{}
\expandafter\ifx\csname natexlab\endcsname\relax\def\natexlab#1{#1}\fi
\providecommand{\url}[1]{\href{#1}{#1}}
\providecommand{\dodoi}[1]{doi:~\href{http://doi.org/#1}{\nolinkurl{#1}}}
\providecommand{\doeprint}[1]{\href{http://ascl.net/#1}{\nolinkurl{http://ascl.net/#1}}}
\providecommand{\doarXiv}[1]{\href{https://arxiv.org/abs/#1}{\nolinkurl{https://arxiv.org/abs/#1}}}

\bibitem[{A. Abeysekara {et~al.}(2017)Abeysekara, Albert, Alfaro, Alvarez,
  {\'A}lvarez, Arceo, Arteaga-Vel{\'a}zquez, Solares, Barber, Bautista-Elivar,
  {et~al.}}]{abeysekara2017observation}
Abeysekara, A., Albert, A., Alfaro, R., {et~al.} 2017,
  \bibinfo{title}{Observation of the crab nebula with the HAWC gamma-ray
  observatory,} The Astrophysical Journal, 843, 39

\bibitem[{A. Abeysekara {et~al.}(2019)Abeysekara, Albert, Alfaro, Alvarez,
  {\'A}lvarez, Camacho, Arceo, Arteaga-Vel{\'a}zquez, Arunbabu, Rojas,
  {et~al.}}]{abeysekara2019measurement}
Abeysekara, A., Albert, A., Alfaro, R., {et~al.} 2019,
  \bibinfo{title}{Measurement of the Crab Nebula spectrum past 100 TeV with
  HAWC,} The Astrophysical Journal, 881, 134

\bibitem[{A. Abeysekara {et~al.}(2023)Abeysekara, Albert, Alfaro, Alvarez,
  {\'A}lvarez, Araya, Arteaga-Vel{\'a}zquez, Arunbabu, Rojas, Solares,
  {et~al.}}]{abeysekara2023high}
Abeysekara, A., Albert, A., Alfaro, R., {et~al.} 2023, \bibinfo{title}{The
  high-altitude water Cherenkov (HAWC) observatory in M{\'e}xico: The primary
  detector,} Nuclear Instruments and Methods in Physics Research Section A:
  Accelerators, Spectrometers, Detectors and Associated Equipment, 1052, 168253

\bibitem[{F. Aharonian {et~al.}(2021)Aharonian, An, Bai, Bai, Bao, Bastieri,
  Bi, Bi, Cai, Cai, {et~al.}}]{aharonian2021observation}
Aharonian, F., An, Q., Bai, L., {et~al.} 2021, \bibinfo{title}{Observation of
  the Crab Nebula with LHAASO-KM2A- a performance study,} Chinese Physics C,
  45, 025002

\bibitem[{A. Albert {et~al.}(2024{\natexlab{a}})Albert, Alfaro, Alvarez,
  Andr{\'e}s, Arteaga-Vel{\'a}zquez, Rojas, Solares, Babu, Belmont-Moreno,
  Bernal, {et~al.}}]{albert2024performance}
Albert, A., Alfaro, R., Alvarez, C., {et~al.} 2024{\natexlab{a}},
  \bibinfo{title}{Performance of the HAWC Observatory and TeV gamma-ray
  measurements of the Crab Nebula with improved extensive air shower
  reconstruction algorithms,} The Astrophysical Journal, 972, 144

\bibitem[{A. Albert {et~al.}(2024{\natexlab{b}})Albert, Alfaro, Alvarez,
  Andr{\'e}s, Arteaga-Vel{\'a}zquez, Rojas, Solares, Babu, Belmont-Moreno,
  Bernal, {et~al.}}]{albert2024observation}
Albert, A., Alfaro, R., Alvarez, C., {et~al.} 2024{\natexlab{b}},
  \bibinfo{title}{Observation of the Galactic Center PeVatron beyond 100 TeV
  with HAWC,} The Astrophysical Journal Letters, 973, L34

\bibitem[{J. Albert {et~al.}(2008)Albert, Aliu, Anderhub, Antoranz, Armada,
  Asensio, Baixeras, Barrio, Bartko, Bastieri,
  {et~al.}}]{albert2008implementation}
Albert, J., Aliu, E., Anderhub, H., {et~al.} 2008,
  \bibinfo{title}{Implementation of the random forest method for the imaging
  atmospheric Cherenkov telescope MAGIC,} Nuclear Instruments and Methods in
  Physics Research Section A: Accelerators, Spectrometers, Detectors and
  Associated Equipment, 588, 424

\bibitem[{R. Alfaro {et~al.}(2022)Alfaro, Alvarez, {\'A}lvarez, Camacho,
  Arteaga-Vel{\'a}zquez, Rojas, Solares, Babu, Belmont-Moreno, Brisbois,
  {et~al.}}]{alfaro2022gamma}
Alfaro, R., Alvarez, C., {\'A}lvarez, J., {et~al.} 2022,
  \bibinfo{title}{Gamma/hadron separation with the HAWC observatory,} Nuclear
  Instruments and Methods in Physics Research Section A: Accelerators,
  Spectrometers, Detectors and Associated Equipment, 1039, 166984

\bibitem[{R. Alfaro {et~al.}(2024)Alfaro, Alvarez, Arteaga-Vel{\'a}zquez,
  Rojas, Solares, Babu, Belmont-Moreno, Bernal, Caballero-Mora, Capistr{\'a}n,
  {et~al.}}]{alfaro2024testing}
Alfaro, R., Alvarez, C., Arteaga-Vel{\'a}zquez, J., {et~al.} 2024,
  \bibinfo{title}{Testing the molecular cloud paradigm for ultra-high-energy
  gamma ray emission from the direction of SNR G106. 3+ 2.7,} Astronomy \&
  Astrophysics, 691, A89

\bibitem[{J. Allison(2007)Allison}]{allison2007facilities}
Allison, J. 2007, \bibinfo{title}{Facilities and methods: Geant4--a simulation
  toolkit,} Nuclear Physics News, 17, 20

\bibitem[{R. Atkins {et~al.}(2003)Atkins, Benbow, Berley, Blaufuss, Bussons,
  Coyne, Delay, DeYoung, Dingus, Dorfan, {et~al.}}]{atkins2003observation}
Atkins, R., Benbow, W., Berley, D., {et~al.} 2003, \bibinfo{title}{Observation
  of tev gamma rays from the crab nebula with milagro using a new background
  rejection technique,} The Astrophysical Journal, 595, 803

\bibitem[{X. Bai {et~al.}(2019)Bai, Bi, Bi, Cao, Chen, Chen, Chiavassa, Cui,
  Dai, Della~Volpe, {et~al.}}]{bai2019large}
Bai, X., Bi, B., Bi, X., {et~al.} 2019, \bibinfo{title}{The Large High Altitude
  Air Shower Observatory (LHAASO) Science White Paper--2021 Edition,} arXiv
  e-prints, arXiv

\bibitem[{R. Brun \& F. Rademakers(1997--2024)Brun \& Rademakers}]{root}
Brun, R., \& Rademakers, F. 1997--2024, \bibinfo{title}{ROOT - An Object
  Oriented Data Analysis Framework,}, \\url{https://root.cern/}

\bibitem[{L. Collaboration*† {et~al.}(2021)Collaboration*†, Cao, Aharonian,
  An, Axikegu, Bai, Bai, Bao, Bastieri, Bi, {et~al.}}]{lhaaso2021peta}
Collaboration*†, L., Cao, Z., Aharonian, F., {et~al.} 2021,
  \bibinfo{title}{Peta--electron volt gamma-ray emission from the Crab Nebula,}
  Science, 373, 425

\bibitem[{T. Fawcett(2006)Fawcett}]{fawcett2006introduction}
Fawcett, T. 2006, \bibinfo{title}{An introduction to ROC analysis,} Pattern
  recognition letters, 27, 861

\bibitem[{D. Heck {et~al.}(1998)Heck, Knapp, Capdevielle, Schatz, Thouw,
  {et~al.}}]{heck1998corsika}
Heck, D., Knapp, J., Capdevielle, J., {et~al.} 1998, \bibinfo{title}{CORSIKA: A
  Monte Carlo code to simulate extensive air showers,} Report fzka, 6019

\bibitem[{A. H{\"o}cker {et~al.}(2008)H{\"o}cker, Speckmayer, Tegenfeldt,
  Stelzer, \& Voss}]{hocker2008tmva}
H{\"o}cker, A., Speckmayer, P., Tegenfeldt, F., Stelzer, J., \& Voss, H. 2008,
  \bibinfo{title}{TMVA: The Toolkit for Multivariate Data Analysis eith ROOT,}

\bibitem[{K. Kamata \& J. Nishimura(1958)Kamata \&
  Nishimura}]{kamata1958lateral}
Kamata, K., \& Nishimura, J. 1958, \bibinfo{title}{The lateral and the angular
  structure functions of electron showers,} Progress of Theoretical Physics
  Supplement, 6, 93

\bibitem[{D.~P. Kingma \& J. Ba(2014)Kingma \& Ba}]{kingma2014adam}
Kingma, D.~P., \& Ba, J. 2014, \bibinfo{title}{Adam: A method for stochastic
  optimization,} arXiv preprint arXiv:1412.6980

\bibitem[{A. Kryukov {et~al.}(2025)Kryukov, Demichev, \&
  Ilyin}]{kryukov2025machine}
Kryukov, A., Demichev, A., \& Ilyin, V. 2025, \bibinfo{title}{Machine Learning
  in Gamma Astronomy,} arXiv preprint arXiv:2501.19064

\bibitem[{J. Li {et~al.}(2025)Li, Lv, Liu, Huang, Wang, \&
  Lin}]{li2025application}
Li, J., Lv, H., Liu, Y., {et~al.} 2025, \bibinfo{title}{Application of Machine
  Learning to Background Rejection in Very-high-energy Gamma-Ray Observation,}
  The Astrophysical Journal Supplement Series, 276, 24

\bibitem[{V. Marandon {et~al.}(2019)Marandon, Jardin-Blicq, \&
  Schoorlemmer}]{marandon2019latest}
Marandon, V., Jardin-Blicq, A., \& Schoorlemmer, H. 2019,
  \bibinfo{title}{Latest news from the HAWC outrigger array,} arXiv preprint
  arXiv:1908.07634

\bibitem[{S. Ohm {et~al.}(2009)Ohm, van Eldik, \& Egberts}]{ohm2009gamma}
Ohm, S., van Eldik, C., \& Egberts, K. 2009, \bibinfo{title}{$\gamma$/hadron
  separation in very-high-energy $\gamma$-ray astronomy using a multivariate
  analysis method,} Astroparticle Physics, 31, 383

\bibitem[{S. Ohm {et~al.}(2023)Ohm, Wagner, Collaboration,
  {et~al.}}]{ohm2023current}
Ohm, S., Wagner, S., Collaboration, H., {et~al.} 2023, \bibinfo{title}{Current
  status and operation of the HESS array of imaging atmospheric Cherenkov
  telescopes,} Nuclear Instruments and Methods in Physics Research Section A:
  Accelerators, Spectrometers, Detectors and Associated Equipment, 1055, 168442

\bibitem[{A. Pagliaro {et~al.}(2011)Pagliaro, Staiti, \&
  D'Anna}]{pagliaro2011discrimination}
Pagliaro, A., Staiti, G.~D., \& D'Anna, F. 2011, \bibinfo{title}{A
  discrimination technique for extensive air showers based on multiscale,
  lacunarity and neural network analysis,} Nuclear Physics B-Proceedings
  Supplements, 212, 286

\bibitem[{A. Paszke {et~al.}(2019)Paszke, Gross, Massa, Lerer, Bradbury,
  Chanan, Killeen, Lin, Gimelshein, Antiga, {et~al.}}]{paszke2019pytorch}
Paszke, A., Gross, S., Massa, F., {et~al.} 2019, in Advances in neural
  information processing systems, 8024--8035

\bibitem[{Z. Tian {et~al.}(2018)Tian, Wang, Liu, Guo, Ma, \&
  Hu}]{tian2018study}
Tian, Z., Wang, Z., Liu, Y., {et~al.} 2018, \bibinfo{title}{Study of the
  $\gamma$/p discrimination at~ 100 TeV energy range with LHAASO experiment,}
  Astroparticle Physics, 99, 43

\bibitem[{X. Wang(2019)Wang}]{wang2019gamma}
Wang, X. 2019, in 36th International Cosmic Ray Conference (ICRC2019), Vol.~36,
  820

\bibitem[{S. Westerhoff {et~al.}(1995)Westerhoff, Funk, Lindner, Magnussen,
  Meyer, M{\"o}ller, Rhode, Sooth, \& Wiebel-Sooth}]{westerhoff1995separating}
Westerhoff, S., Funk, B., Lindner, A., {et~al.} 1995,
  \bibinfo{title}{Separating $\gamma$-and hadron-induced cosmic ray air showers
  with feed-forward neural networks using the charged particle information,}
  Astroparticle Physics, 4, 119

\end{thebibliography}
\bibliographystyle{aasjournalv7}

\end{document}